\documentclass[floats,prb,aps,twocolumn]{revtex4}
\usepackage{graphicx}
\begin{document}

\title{Phase transitions in a system of indirect magnetoexcitons  in
coupled quantum wells
  at high magnetic field: the role of disorder}

\author{Oleg L. Berman$^{1}$, Yurii E. Lozovik$^{2}$,  David W.
Snoke$^{3}$, and Rob D. Coalson$^{1}$}

\affiliation{\mbox{$^{1}$ Department of Chemistry, University of
Pittsburgh,}   \\ Pittsburgh, PA 15260, USA  \\
\mbox{$^{2}$ Institute of Spectroscopy, Russian Academy of
Sciences,}  \\ 142190 Troitsk, Moscow Region, Russia \\
\mbox{$^{3}$Department of Physics and Astronomy, University of
Pittsburgh,}  \\ Pittsburgh, PA 15260 USA  }


\begin{abstract}
Collective properties of a quasi-two-dimensional ($2D$) system of
spatially indirect magnetoexcitons in coupled quantum wells (CQW)
in high magnetic field $H$ were analyzed in the presence of
  disorder. The Hamiltonian of the dilute gas of magnetoexcitons with
dipole-dipole repulsion in a random field has been reduced to the
Hamiltonian of a dilute gas of dipolar excitons without an applied
magnetic field, but in an $H$-dependent effective random field and
  having an effective mass of magnetoexciton which is
  a function of the magnetic field and parameters of the CQW.
   For $2D$ magnetoexcitonic systems, the increase of the magnetic field $H$
and the interwell distance $D$ is found to increase  the effective
renormalized random field parameter $Q$ and suppress the
superfluid density $n_s$ and the temperature of the
Kosterlitz-Thouless transition $T_c$. It is shown that in the
presence of the disorder there is a quantum transition to the
superfluid state at zero temperature $T=0$ with respect to the
magnetic field $H$ and the parameters of the disorder. There is no
superfluidity at any exciton density in the presence of the
disorder at sufficiently large magnetic field $H$ or sufficiently large
 disorder.

\vspace{0.1cm}



\end{abstract}

\maketitle

Indirect excitons  in coupled  quantum wells (CQW) both with and
without magnetic fields ($H$) have been the subject of recent
experimental investigations  (see Fig.1).
\cite{Chemla}$^-$\cite{Butov_Chemla} These systems are of
interest, in particular, in connection with the possibility of
Bose-Einstein condensation and superfluidity of indirect excitons
or electron-hole pairs, which would manifest itself in the CQW as
persistent electrical currents in each well and also through
coherent optical properties and Josephson phenomena.
\cite{Lozovik}$^-$\cite{Berman_Willander} In high magnetic fields
($H > 7 $T) two-dimensional (2D)
  excitons survive in a substantially wider temperature region, as the
exciton binding energies increase with magnetic field.
\cite{Lerner}$^-$\cite{Moskalenko} The theory of magnetoexcitonic
systems developed to date has not taken into account the influence
of random field on the phase transitions, which is created by
impurities and boundary irregularities of the quantum wells. In
real experiments, however, disorder plays a very important role.
Although the inhomogeneous broadening linewidth of typical
GaAs-based samples has been improved from around 20 meV to less
than 1 meV,\cite{Snoke_paper} this disorder energy is still not
much smaller than the exciton-exciton repulsion energy. (At a
typical exciton density of $10^{10}$ cm$^{-2}$, the interaction
energy of the excitons is approximately $1$ meV.\cite{Snoke})  On
the other hand, the typical disorder energy of 1 meV is low
compared to the typical exciton binding energy of 5 meV.

In the present Letter we study superfluidity in a ``dirty'' system
of indirect excitons in strong magnetic field $H$. We reduce the
problem of magnetoexcitons in random fields to the problem of
  excitons at $H = 0$ and in a renormalized random field depending on
  $H$. We analyze the dependence of Kosterlitz-Thouless
transition\cite{Kosterlitz}
temperature
  and superfluid density on magnetic field.

The total Hamiltonian $\hat H$ describing 2D spatially separated
electrons ($e$) and holes ($h$) in a perpendicular magnetic field
in the presence of the external field has the form:
\begin{eqnarray}\label{H_Tot}
\hat H &=& \int d \mathbf{R} \int d\mathbf{r}
\left[\hat{\psi}^{\dagger}(\mathbf{R},\mathbf{r}) \right.
  \nonumber \\ && \left. \left(\frac{1}{2m_{e}} \left( -i
\nabla_{e} + e\mathbf{A}_{e} \right)^2 + \frac{1}{2m_{h}}\left( -i
\nabla_{h} - e\mathbf{A}_{h} \right)^2 \right.\right. \nonumber \\
&-& \left.\left.\frac{e^2}{\epsilon \sqrt{(\mathbf{r}_{e} -
\mathbf{r}_{h})^{2} + D^{2}}} + V_{e}(\mathbf{r}_{e}) +
V_{h}(\mathbf{r}_{h})\right)\hat{\psi}(\mathbf{R},\mathbf{r})\right]
\nonumber \\ &+& \frac{1}{2}\int d \mathbf{R}_{1} \int
d\mathbf{r}_{1}\int d \mathbf{R}_{2} \int d\mathbf{r}_{2}
\hat{\psi}^{\dagger}(\mathbf{R}_{1},\mathbf{r}_{1})\hat{\psi}^{\dagger}(\mathbf{R}_{2},\mathbf{r}_{2})
\nonumber \\ && \left(U^{ee}(\mathbf{r}_{e1} - \mathbf{r}_{e2}) +
  U^{hh}(\mathbf{r}_{h1} -
\mathbf{r}_{h2}) + U^{eh}(\mathbf{r}_{e1} - \mathbf{r}_{h2})
\right. \nonumber \\ \  &+& \left.  U^{he}(\mathbf{r}_{h1} -
\mathbf{r}_{e2})
\right)\hat{\psi}(\mathbf{R}_{2},\mathbf{r}_{2})\hat{\psi}(\mathbf{R}_{1},\mathbf{r}_{1}).
\end{eqnarray}
Here $\hat{\psi}^{\dagger}(\mathbf{R},\mathbf{r})$ and
$\hat{\psi}(\mathbf{R},\mathbf{r})$ are the creation and
annihilation operators for magnetoexcitons;
  $\mathbf{r}_{e}$ and $\mathbf{r}_{h}$ are electron and hole
locations along quantum wells, correspondingly; $\mathbf{A}_{e}$,
$\mathbf{A}_{h}$ are the vector potentials at the electron and
hole location, respectively; $V_{e}(\mathbf{r}_{e})$ and
$V_{h}(\mathbf{r}_{h})$ represent the external fields acting on
electron and hole, respectively (we use units $c = \hbar = 1$);
  $D$ is the
distance between electron and hole quantum wells; $e$ is the
charge of an electron; $\epsilon$ is the dielectric constant. We
use below the coordinates of the magnetoexciton center of mass
$\mathbf{R} = (m_{e}\mathbf{r}_{e} + m_{h}\mathbf{r}_{h})/(m_{e} +
m_{h})$
  and the internal exciton coordinates
  $\mathbf{r} = \mathbf{r}_{e} - \mathbf{r}_{h}$.
The cylindrical gauge for vector-potential is used:
$\mathbf{A}_{e,h} = \frac{1}{2}\mathbf{H}\times \mathbf{r}_{e,h}$.
  $U^{ee}$, $U^{hh}$,
$U^{eh}$ and $U^{he}$ are the two-particle potentials of the
electron-electron, hole-hole, electron-hole and hole-electron
interaction, respectively, between electrons or holes from
different pairs: $U^{ee}(\mathbf{r}_{e1} - \mathbf{r}_{e2}) =
e^{2}/(\epsilon |\mathbf{r}_{e1} - \mathbf{r}_{e2}|)$; $
U^{hh}(\mathbf{r}_{h1} - \mathbf{r}_{h2}) = e^{2}/(\epsilon
|\mathbf{r}_{h1} - \mathbf{r}_{h2}|)$; $U^{eh}(\mathbf{r}_{e1} -
\mathbf{r}_{h2}) = - e^{2}/(\epsilon \sqrt{|\mathbf{r}_{e1} -
\mathbf{r}_{h2}|^{2} + D^{2}})$; $U^{he}(\mathbf{r}_{h1} -
\mathbf{r}_{e2}) = - e^{2}/(\epsilon \sqrt{|\mathbf{r}_{h1} -
\mathbf{r}_{e2}|^{2} + D^{2}})$.

Note that the exciton {\it magnetic} momentum \cite{GorDzyal}
$\hat{\mathbf{P}} = -i \nabla _{e} -i \nabla _{h} + e
(\mathbf{A}_{e} - \mathbf{A}_{h}) - e\mathbf{ H} \times
(\mathbf{r}_{e} - \mathbf{r}_{h})$ is  a conserved quantity for an
isolated exciton in a magnetic field without any external field
($V_{e}(\mathbf{r}_{e}) = V_{h}(\mathbf{r}_{h}) = 0$). The
eigenfunctions of the Hamiltonian of a single isolated magnetoexciton
without any random field ($V_{e}(\mathbf{r}_{e}) =
V_{h}(\mathbf{r}_{h}) = 0$), which are also the eigenfunctions of
the magnetic momentum $\bf{P}$, have the following form (see
Refs.~[\onlinecite{Lerner,GorDzyal}]):
\begin{eqnarray}\label{W_Func_Gen}
&&\Psi _{k\mathbf{P}} ({\bf R},{\bf r}) = \nonumber\\ && \exp \left\{i {\bf R}
\left( {\bf P} + \frac{e}{2} {\bf H}\times {\bf R} \right) + i
\gamma \frac{\mathbf{P}\mathbf{r}}{2} \right\} \Phi_{k}
(\mathbf{P},\mathbf{r}) ,
\end{eqnarray}
where $\Phi _{k} ({\bf P},{\bf r})$ is a function of internal
coordinates ${\bf r}$; ${\bf P}$ is the eigenvalue of magnetic
momentum; $k$ represents the quantum numbers of exciton internal
motion. In high magnetic fields
  $k = (n_{L},m)$, where  $n_{L} = min(n_{1}, n_{2})$, $m = |n_{1} -
n_{2}|$, and $n_{1(2)}$ are Landau
  quantum numbers for $e$ and $h$
\cite{Lerner}$^,$\cite{Ruvinskiy}; $\gamma = (m_h - m_e)/(m_h +
m_e)$.

We further expand the magnetoexciton field operators in a single
magnetoexciton basis set $\Psi _{k\mathbf{P}} ({\bf R},{\bf r})$:
$\hat{\psi}^{\dagger}(\mathbf{R},\mathbf{r}) = \sum_{k\mathbf{P}}
\Psi _{k\mathbf{P}}^{*} ({\bf R},{\bf
r})\hat{a}_{k\mathbf{P}}^{\dagger}$;
$\hat{\psi}(\mathbf{R},\mathbf{r}) = \sum_{k\mathbf{P}} \Psi
_{k\mathbf{P}}({\bf R},{\bf r})\hat{a}_{k\mathbf{P}}$,  where
$\hat{a}_{k\mathbf{P}}^{\dagger}$ and $\hat{a}_{k\mathbf{P}}$ are
the corresponding  creation and annihilation operators of a
magnetoexciton in $(k,\mathbf{P})$ space.

We consider the case of strong magnetic field, when we neglect in
Eq.~(\ref{W_Func_Gen}) the transitions between different Landau
levels of the magnetoexciton caused by scattering by the slowly
changing in space potential $V_{e}(\mathbf{r}_{e}) +
V_{h}(\mathbf{r}_{h})$. We also neglect nondiagonal  matrix
elements of the Coulomb interaction between a paired electron and
hole. The region of applicability of these two assumptions is
defined by the inequalities\cite{Ruvinsky_jetp} $\omega_{c} \gg
E_{b}$, $\omega_{c} \gg \sqrt{\left\langle
V_{e(h)}^{2}\right\rangle _{av}}$, where $\omega_{c} =
eH/m_{e-h}$, $m_{e-h} = m_{e}m_{h}/(m_{e} + m_{h})$ is the exciton
reduced mass in the quantum well plane; $E_{b}$ is the
magnetoexciton binding energy in an ideal ``pure'' system as as a
function of magnetic field $H$ and the distance between electron
and hole quantum wells $D$: $E_{b} \sim e^{2}/\epsilon
r_{H}\sqrt{\pi/2}$ at $D \ll r_{H}$ and $E_{b} \sim e^{2}/\epsilon
D$ at $D \gg r_{H}$ ($r_{H} = (eH)^{-1/2}$ is the magnetic
length).\cite{Lerner,Ruvinskiy} Here $\left\langle \ldots
\right\rangle _{av}$ denotes averaging over the fluctuations of
the random field.

In a strong magnetic field at low densities (~$n \ll r_{H}^{-2}$~)
indirect magnetoexcitons repel as  parallel
dipoles,\cite{Berman_Tsvetus}, and we have for the pair
interaction potential:
\begin{eqnarray}\label{dipole_s}
\hat{U}(|\mathbf{R}_{1} - \mathbf{R}_{2}|) \equiv \hat{U}^{ee} +
\hat{U}^{hh} +  \hat{U}^{eh} + \hat{U}^{he} \simeq
\frac{e^{2}D^{2}}{\epsilon|\mathbf{R}_{1} - \mathbf{R}_{2}|^{3}}.
\end{eqnarray}

Now we substitute the expansions  for the field creation and
annihilation operators into the total Hamiltonian
Eq.~(\ref{H_Tot}) and obtain the effective Hamiltonian in terms of
creation and annihilation operators in $\mathbf{P}$ space.
  In high
magnetic field, when the typical interexciton interaction
$D^{2}n^{-\frac{3}{2}} \ll \omega _{c}$,
  one can ignore transitions between Landau levels and consider
only the states corresponding to the lowest Landau level
$m=n_{L}=0$. Using the orthonormality of the functions $\Phi _{mn}
({\bf 0},\mathbf{r})$ we obtain the effective Hamiltonian
$\hat{H}_{\rm eff}$ in strong magnetic fields. Since a typical value
of $r$ is $r_H$, and $P \ll 1/r_H$, in this approximation the
effective Hamiltonian $\hat{H}_{\rm eff}$ in the magnetic momentum
representation $P$ in the subspace the lowest Landau level
$m=n_{L}=0$ has the same form  (compare with
Ref.[\onlinecite{Berman_Snoke_Coalson}])
  as for two-dimensional boson
system without a magnetic field, but with  the magnetoexciton
magnetic mass $m_{H}$ (which depends on $H$ and $D$; see below)
instead of the exciton mass ($M = m_{e} + m_{h}$), magnetic
momenta instead of ordinary momenta and renormalized random field
(for the lowest Landau level we denote the spectrum of the single
exciton $\varepsilon _{0}(P) \equiv \varepsilon _{00}({\bf P})$):
\begin{eqnarray}\label{H_eff}
&& \hat{H}_{\rm eff} = \sum_{\mathbf{P}}  \varepsilon_{0}(P)
\hat{a}_{\mathbf{P}}^{\dagger}\hat{a}_{\mathbf{P}}  +
\sum_{\mathbf{P},\mathbf{P}'}\left\langle
\mathbf{P}'\left|\hat{V}\right|\mathbf{P}\right\rangle
\hat{a}_{\mathbf{P}'}^{\dagger} \hat{a}_{\mathbf{P}} +
\frac{1}{2}\nonumber
\\  &&
\sum_{\mathbf{P}_{1},\mathbf{P}_{2},\mathbf{P}_{3},\mathbf{P}_{4}}
  \left\langle
\mathbf{P}_{1},\mathbf{P}_{2}\left|\hat{U}\right|\mathbf{P}_{3},\mathbf{P}_{4}\right\rangle
\hat{a}_{\mathbf{P}_{1}}^{\dagger}
\hat{a}_{\mathbf{P}_{2}}^{\dagger}\hat{a}_{\mathbf{P}_{3}}\hat{a}_{\mathbf{P}_{4}},
\end{eqnarray}
where $\hat{V} = \hat{V}_{e} + \hat{V}_{h}$. The dispersion
relation $\varepsilon _{0}(P)$ of an isolated magnetoexciton  on
the lowest Landau level is a quadratic function at the small
magnetic momenta under consideration, $\varepsilon _{0}({\bf P})
\approx P^2/(2m_{H })$, where $m_{H }$ is the effective {\it
magnetic} mass of a magnetoexciton in the lowest Landau level,
dependent on $H$ and the distance $D$ between $e$ -- and $h$ --
layers (see Ref.[\onlinecite{Ruvinskiy}]). In strong magnetic
fields at $D \gg r_{H}$ the exciton magnetic mass is $m_H \approx
D^{3}\epsilon/(e^{2}r_{H}^{4})$ \cite{Ruvinskiy}.

The matrix element $\left\langle
\mathbf{P}_{1},\mathbf{P}_{2}\left|\hat{U}\right|\mathbf{P}_{3},\mathbf{P}_{4}\right\rangle$
is the Fourier transform of the pair interaction potential
(Eq.~(\ref{dipole_s})). In the strong magnetic field limit, using
for $\Phi _{k} ({\bf 0},\mathbf{r})$ the internal wavefunction of
the magnetoexciton at the lowest Landau level $\Phi _{k=0} ({\bf
0},\mathbf{r})$, \cite{Ruvinskiy} we obtain the matrix element of
the external potential $V_{e,h}(\mathbf{r})$ connecting the states
$\left\langle k=0, \mathbf{P}\left|\right.\right.$ and
$\left\langle k=0,\mathbf{P}'\left|\right.\right.$:
\begin{eqnarray}
\label{mat_ext} && \left \langle{\bf P}'\mid
\hat{V}_{e,h}(\mathbf{r}) \mid {\bf P}\right \rangle=\frac{1}{S}
\exp\left(-({\bf P}'-{\bf P})^2\frac{r_{H}^2}{4}\right) \nonumber
\\ &\times & V_{e,h}({\bf P}'-{\bf P}) \exp\left(\pm
\frac{ir_{H}^2}{2H}{\bf H}\cdot ({\bf P}\times{\bf P}')\right),
\end{eqnarray}
where $V_{e,h}({\bf P}'-{\bf P})$ is the Fourier transform of
$V_{e,h} (\mathbf{r})$.

Thus, the effective Hamiltonian $\hat{H}_{\rm eff}$ Eq.~(\ref{H_eff})
corresponds to excitons with renormalized dispersion law
$\varepsilon_{0}(P)$  and effective random field
\begin{eqnarray}\label{V_eff}
V_{\rm eff}(\mathbf{R}) &=& \frac{1}{\pi r_{H}^{2}} \int \exp\left(
-\frac{(\mathbf{R}- \mathbf{r})^{2}}{r_{H}^{2}}\right) \nonumber
\\ && \left[ V_{e}(\mathbf{r}) + V_{h}(\mathbf{r}) \right] d \mathbf{r}.
\end{eqnarray}
The approach is valid, if the correlation length $L$ of random
potential ($V(\mathbf{r}_{e},\mathbf{r}_{h})$) is much greater
than the magnetoexciton mean size\cite{Ruvinsky_jetp} $r_{exc} =
r_{H}$: ($L \gg r_{H}$) or $r_{exc}\sqrt{\left\langle\nabla
V^{2}\right\rangle_{av}} \ll E_{b}$, and it holds for the strong
magnetic field, when $r_{exc} = r_{H} =(eH)^{-1/2}$, and $E_{b}
\sim e^{2}/\epsilon D$ at $D \gg
r_{H}$\cite{Ruvinskiy,Ruvinsky_jetp}.

The interaction between an spatially indirect exciton in coupled
quantum wells and a random field, induced by fluctuations in
widths of electron and hole quantum wells, has the form
\cite{Ruvinsky_jetp} $ V(\mathbf{r}_{e},\mathbf{r}_{h}) =
\alpha_{e}[\xi_{1}(\mathbf{r}_{e}) - \xi_{2}(\mathbf{r}_{e})] +
\alpha_{h}[\xi_{3}(\mathbf{r}_{h}) - \xi_{4}(\mathbf{r}_{h})]$,
where $\alpha_{e,h} = \partial E_{e,h}^{(0)}/\partial d_{e,h}$,
$d_{e,h}$ is the average  widths of the electron,hole quantum
wells, while $E_{e,h}^{(0)}$ are the lowest levels of the electron
and hole in the conduction and valence bands, and $\xi_{1}$ and
$\xi_{2}$ ($\xi_{3}$ and $\xi_{4}$) are fluctuations in the widths
of the electron (hole) wells on the upper and lower interfaces,
respectively. We assume that fluctuations on different interfaces
are statistically independent, whereas fluctuations of a specific
interface are characterized by Gaussian correlation function $
\left\langle \xi_{i}(\mathbf{r}_{1})\xi_{j}(\mathbf{r}_{2})
\right\rangle = g_{i}\delta_{ij}\delta(\mathbf{r}_{2} -
\mathbf{r}_{1})$, $ \left\langle \xi_{i}(\mathbf{r}) \right\rangle
= 0$, where $g_{i}$ is proportional to the squared amplitude of
the $i$th interface fluctuation\cite{Ruvinsky_jetp}. This is
possible if the distance $D$ between the electron and hole quantum
wells is larger than the amplitude of fluctuations on the nearest
surfaces.

We consider the characteristic length of the random field
potential  $L$ to be much shorter than the average distance
between excitons $r_{s} \sim 1/\sqrt{\pi n}$  ($L \ll 1/\sqrt{\pi
n}$, where $n$ is the total exciton density) similar to
Ref.~[\onlinecite{Berman_Snoke_Coalson}]. Since the effective
Hamiltonian  $\hat{H}_{\rm eff}$  of the system of indirect ``dirty''
magnetoexcitons at small momenta  is exactly identical to the
Hamiltonian of indirect ``dirty'' excitons without magnetic field
but with magnetic mass $m_{H}$ and effective random field
$V_{\rm eff}$ instead of  $M = m_{e} + m_{h}$ and $ V_{e}(\mathbf{r})
+ V_{h}(\mathbf{r})$, respectively, we can use the expressions
for the ladder approximation
  Green's function\cite{Yudson}, collective spectrum, normal and superfluid
  density and the temperature of Kosterlitz-Thouless phase
  transition \cite{Kosterlitz} for the ``dirty''
\cite{Berman_Snoke_Coalson} system without magnetic field.

  The spectrum of interacting excitons  has the form (cf.
Ref.~[\onlinecite{Berman_Snoke_Coalson}])
$$
\varepsilon (p) =
\sqrt{\left(p^{2}/(2m_{H}) + \sqrt{\mu^{2} - Q^{2}}\right)^{2} -
(\mu^{2} - Q^{2})},
$$ and for small momenta $p \ll
\sqrt{2m_{H}\mu}$ the excitation spectrum is acoustic $\varepsilon
(p) = c_{s} p$, where $c_{s} = \sqrt{\sqrt{\mu^{2} -
Q^{2}}/m_{H}}$ is the velocity of sound. The chemical potential
$\mu$ in the ladder approximation has the
form\cite{Berman_Snoke_Coalson} $ \mu = 8\pi n/\left[2m_{H} \log
\left(\epsilon^{2}/(8\pi n m_{H}^2 e^4 D^4) \right)\right]$. In
the weak-scattering limit  we use the second-order Born
approximation for the random field parameter $Q$ similar to
Refs.~[\onlinecite{Berman_Snoke_Coalson}], and for small
frequencies and momenta, which mostly contribute to the ladder
approximation Green's function,  approximate $Q (\mathbf{p},
\omega)$ by $Q (\mathbf{p} = \mathbf{0}, \omega = 0)$
\begin{eqnarray}
\label{Q_res_ap}  Q (\mathbf{p}, \omega) &=& Q =
\frac{\alpha_{e}^{2}(g_{1} + g_{2}) + \alpha_{h}^{2}(g_{3} +
g_{4})}{64\pi^{4}}m_{H}.
\end{eqnarray}

  The density of the superfluid
component $n_{s}(T)$ can be obtained using the relation $n_{s}(T)
= n - n_{n}(T)$, where $n_{n}(T)$ is the density of the normal
component. The density of normal component $n_{n}$ is (compare to
  Ref.~[\onlinecite{Berman_Snoke_Coalson}]):
\begin{eqnarray}
\label{nn55} n_{n} = \frac{3 \zeta (3) }{2 \pi }
\frac{T^3}{c_{s}^{4}(n,Q) m_{H}} + \frac{n
Q}{2m_{H}c_{s}^{2}(n,Q)}.
\end{eqnarray}
  From Eq.~(\ref{nn55}) we can see, that the
random field decreases the density of the superfluid component.

In a 2D system, superfluidity appears below the
Kosterlitz-Thouless transition temperature $T_{c} = \pi
n_{s}/(2m_{H})$\cite{Kosterlitz}, where only coupled vortices are
present. Using the expression~(\ref{nn55}) for the density $n_{s}$
of the superfluid component, we obtain an equation for the
Kosterlitz-Thouless transition temperature $T_{c}$. Its solution
is
\begin{eqnarray}
\label{tct} T_c &=& \left[\left( 1 +
\sqrt{\frac{32}{27}\left(\frac{m_{H} T_{c}^{0}}{\pi n'}\right)^{3}
+ 1} \right)^{1/3}   \right.  \nonumber \\ &-& \left.  \left(
\sqrt{\frac{32}{27} \left(\frac{m_{H} T_{c}^{0}}{\pi
n'}\right)^{3} + 1} - 1 \right)^{1/3}\right] \frac{T_{c}^{0}}{
2^{1/3}}   .
\end{eqnarray}
Here $T_{c}^{0}$ is an auxiliary quantity, equal to the
temperature at which the superfluid density vanishes in the
mean-field approximation $n_{s}(T_{c}^{0}) = 0$, $T_c^0 = \left( 2
\pi n' c_s^4 m_{H}/(3 \zeta (3)) \right)^{1/3}$, $n' = n  - n
Q/(2m_{H}c_{s}^{2})$.

Since in strong magnetic fields at $D \gg r_{H}$ the exciton
magnetic mass is $m_H \approx
D^{3}\epsilon/(e^{2}r_{H}^{4})$,\cite{Ruvinskiy} the
superfluid density $n_s$ and the temperature of the
Kosterlitz-Thouless transition $T_c$
decrease with increase of the magnetic field $H$, and the parameters
of the random field
$\alpha_{e}$, $\alpha_{h}$ and $g_{i}$.  Since in the ``dirty''
systems $n_{s}$ and $T_{c}$ decrease with the increase of
effective random field $Q$ (analogous to the case without magnetic
field\cite{Berman_Snoke_Coalson}), and in a strong
magnetic field  $Q$ is proportional to $m_{H}$
(Eq.~(\ref{Q_res_ap})), an increase of the magnetic field $H$
increases  the effective renormalized random field $Q$, and thus
suppresses the superfluid density $n_s$ and the temperature of the
Kosterlitz-Thouless transition $T_c$.

  It follows from Eq.~(\ref{nn55}) that in  the
presence of disorder at $T=0$ there is a quantum transition from
the superfluid state to a Bose glass at a sufficiently large value
  of the magnetic field $H$ and the parameters of the disorder
$\alpha_{e}$, $\alpha_{h}$ and $g_{i}$. While in the ``pure''
system at any magnetic field $H$ at there is always a region in
the density-temperature space, where the superfluidity
occurs\cite{Berman_Tsvetus}, in the presence of the disorder at
sufficiently large magnetic field $H$ or the parameters of the
disorder $\alpha_{e}$, $\alpha_{h}$ and $g_{i}$ there is no
superfluidity at any exciton density.

Note also that in a magnetic field the superfluid density $n_s$
    and the temperature of the Kosterlitz-Thouless transition  $T_c$
    decrease when the separation between quantum wells $D$ increases.
    As $D$ increases, so do $m_{H}$ \cite{Ruvinskiy}  and thus $Q$
(Eq.~(\ref{Q_res_ap})): increasing either
    of these parameters decreases $n_s$  and $T_c$ (Eqs.~(\ref{nn55})
and~(\ref{tct})).  There is a competing
     influence, namely that increasing $D$ increases the velocity of sound
     (since it increases the chemical potential of the dipole-dipole
repulsion $\mu$),
     which tends to increase  $n_s$  and $T_c$, but this is a logarithmic
     dependence.  Thus, in a strong magnetic field the first two
influences dominate, and $n_s$  and $T_c$
        decrease with $D$. In the absence of a magnetic field, the
first two influences are absent: in this case,
     the aforementioned logarithmic dependence dominates, causing
$n_s$  and $T_c$  to increase with $D$.\cite{Berman_Snoke_Coalson}

In the high magnetic field limit at high $D$, the effective random
field is not small, and approaches which assume coupling with the
random field to be much smaller than the dipole-dipole repulsion
are not applicable. Note that in the present work the parameter
$Q/\mu$ is not required to be very small, and our formulas for the
superfluid density and Kosterlitz-Thouless temperature can be used
in the regime of realistic experimental parameters
  taken from photoluminescence line broadening
  measurements\cite{Snoke_paper}.

\begin{center}
{\bf Acknowledgements}
\end{center}

O. L.~B. wishes to thank the participants of the First
International Conference on Spontaneous Coherence in Excitonic
Systems (ICSCE) in Seven Springs PA for many useful and
stimulating discussions. Yu. E.~L. was supported by the INTAS
grant. D. W.~S. and R. D.~C. have been supported by the National
Science Foundation.

\begin{center}
{\bf Caption to Figure 1}

\end{center}

Fig.1 The geometry of the spatially separated electron ($e$) and
hole ($h$) in coupled quantum wells in external magnetic
(${\mathbf H}$) and electric  (${\mathbf E}$) fields normal to
quantum wells.

\pagebreak

\pagebreak

\pagebreak

\newpage

\begin{figure}
\includegraphics[width = 20cm, height = 17cm]{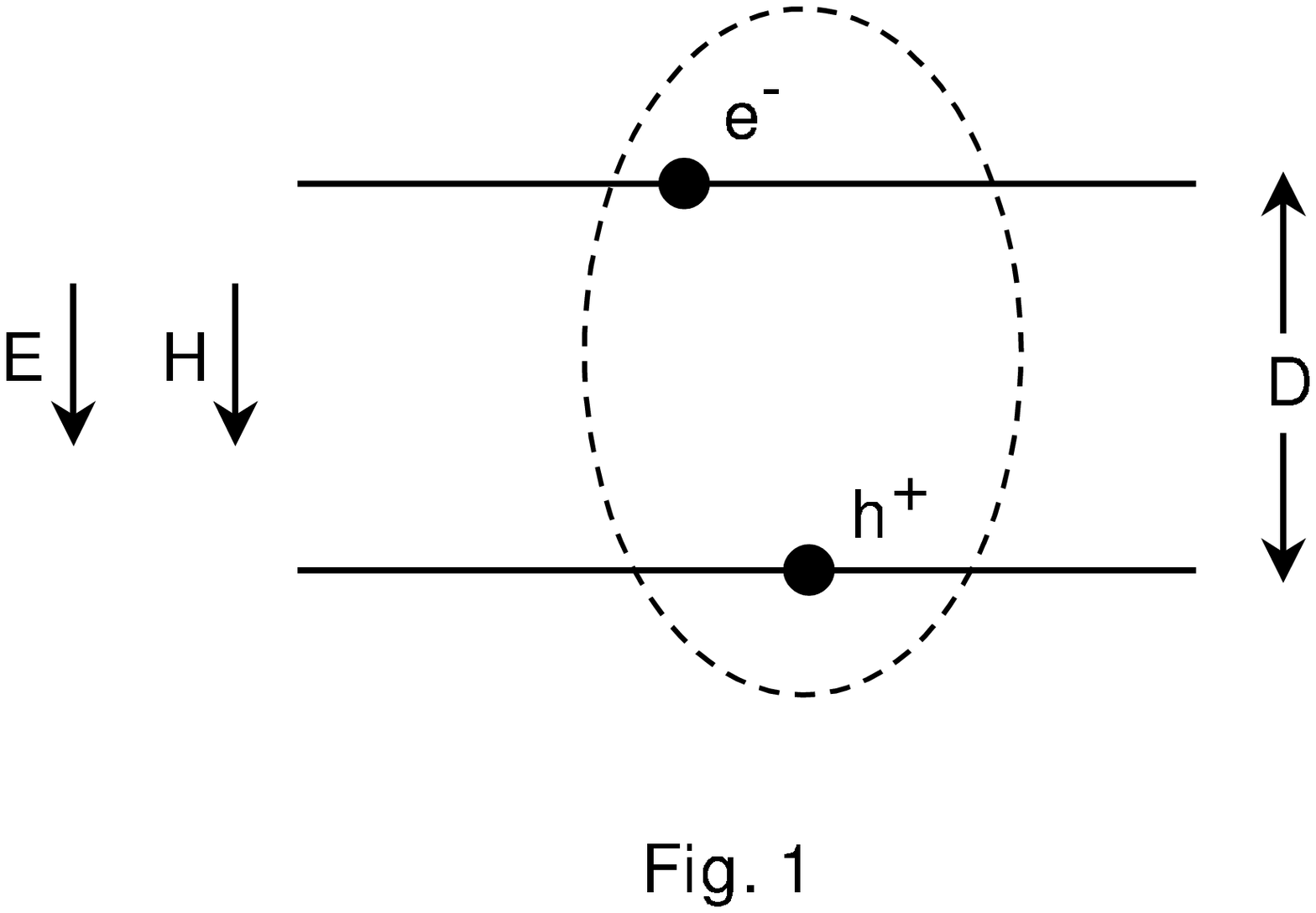}
\end{figure}
\vspace{10cm}
\newpage
\newpage
\pagebreak

\end{document}